

\input harvmac
\def\Title#1#2{\rightline{#1}\ifx\answ\bigans\nopagenumbers\pageno0\vskip1in
\else\pageno1\vskip.8in\fi \centerline{\titlefont #2}\vskip .5in}

%
%
\font\ticp=cmcsc10
\def\sq{{\vbox {\hrule height 0.6pt\hbox{\vrule width 0.6pt\hskip 3pt
   \vbox{\vskip 6pt}\hskip 3pt \vrule width 0.6pt}\hrule height 0.6pt}}}
\def\ajou#1&#2(#3){\ \sl#1\bf#2\rm(19#3)}
\def\frac#1#2{{#1 \over #2}}

\def\eg{{\it e.g.}}
\def\pmi{{\partial_-}}
\def\p+{{\partial_+}}
\def\la{{\lambda}}
\def\ghat{{\hat g}}
%
%
\lref\BuCh{T.T. Burwick and H. Chamseddine, ``Classical and quantum
considerations of two-dimensional gravity,'' Z\"urich preprint ZU-TH-4/92.}
\lref\past{Y. Park and A. Strominger, ``Supersymmetry and Positive
Energy for Classical and Quantum Dilaton Gravity'', to appear.}
\lref\mik{A. Mikovic, ``Exactly solvable
models of 2d dilaton quantum gravity,''Queen Mary preprint QMW/PH/92/12,
hep-th@xxx/9207006.}
\lref\ham{ K. Hamada, ``Quantum theory of dilaton gravity in 1+1 dimensions,''
preprint UT-Komaba 92-7, hep-th@xxx/9206071}
\lref\tbo{T. Banks and M. O'Loughlin, ``Two-dimensional quantum gravity
 in Minkowski space,'' \ajou Nucl. Phys. &B362 (91) 649.}
\lref\jptb{S. R. Das, S. Naik and S. R. Wadia, ``Quantization of the
Liouville
 mode and string theory,'' \ajou Mod.
Phys. Lett. &A4 (89) 1033\semi
J. Polchinski, ``A two-dimensional model for quantum
gravity,'' \ajou Nucl. Phys. &B324 (89) 123\semi
T. Banks and J. Lykken, ``String theory and two-dimensional quantum
gravity,'' \ajou Nucl. Phys. &B331 (90) 173.}
\lref\rut {J. G. Russo and A. A. Tseytlin, {\it Scalar-Tensor Quantum Gravity
in Two Dimensions} preprint SU-ITP-92-2, DAMTP-1-1992.}
\lref\hver{H. Verlinde,
``Black
holes and strings in two dimensions,'' Princeton preprint PUPT-1303,
to appear in the proceedings of the Sixth Marcel Grossman Meeting.}

\lref\RuTs{J.G. Russo and A.A. Tseytlin, ``Scalar-tensor quantum gravity
in two dimensions,'' Stanford/Cambridge preprint SU-ITP-92-2=DAMTP-1-1992.}
\lref\Mandal{G. Mandal, A Sengupta, and S. Wadia, ``Classical solutions of
two-dimensional
string theory,'' \ajou Mod. Phys. Lett. &A6 (91) 1685.}
\lref\WittTwod{E. Witten, ``On string theory and black holes,''\ajou Phys. Rev&
D44 (91) 314.}
\lref\CGHS{C.G. Callan, S.B. Giddings, J.A. Harvey, and A. Strominger,
``Evanescent black holes,"\ajou Phys. Rev. &D45 (92) R1005.}
\lref\BDDO{T. Banks, A. Dabholkar, M.R. Douglas, and M O'Loughlin, ``Are
horned particles the climax of Hawking evaporation?'' \ajou Phys. Rev.
&D45 (92) 3607.}
\lref\rst{J.G. Russo, L. Susskind, and L. Thorlacius, ``The
Endpoint of Hawking Evaporation,'' Stanford preprint SU-ITP-92-17.}
\lref\RST{J.G. Russo, L. Susskind, and L. Thorlacius, ``Black hole
evaporation in 1+1 dimensions,'' Stanford preprint SU-ITP-92-4.}
\lref\Stro{A. Strominger, ``Fadeev-Popov ghosts and 1+1 dimensional black
hole evaporation,'' UCSB preprint UCSBTH-92-18, hep-th@xxx/9205028.}
\lref\deAli{S.P. deAlwis, ``Quantization of a theory of 2d dilaton
gravity,'' Boulder preprint COLO-HEP-280, hep-th@xxx/9205069.}
\lref\deAlii{S.P. deAlwis, ``Black hole physics from Liouville theory,''
Boulder preprint COLO-HEP-284, hep-th@xxx/9206020.}
\lref\BiCa{A. Bilal and C. Callan, ``Liouville models of black hole
evaporation,'' Princeton preprint PUPT-1320, hep-th@xxx/9205089.}
\lref\GiStunpub{S.B. Giddings and A. Strominger, unpublished.}
\lref\CrFu{S. M. Christensen and S. A. Fulling, ``Trace anomalies and the
Hawking effect,''\ajou Phys. Rev. &D15 (77) 2088.}
\lref\GiNe{S.B. Giddings and W.M. Nelson, ``Quantum emission from
two-dimensional black holes,'' UCSB preprint UCSBTH-92-15,
hep-th@xxx/9204072.}
\lref\Hawk{S.W. Hawking, ``Evaporation of two dimensional black holes,''
CalTech preprint CALT-68-1774, hep-th@xxx/9203052.}
\lref\SuTh{L. Susskind and L. Thorlacious, ``Hawking radiation and
back-reaction,'' Stanford preprint SU-ITP-92-12, hep-th@xxx/9203054.}
\lref\BGHS{B. Birnir, S.B. Giddings, J.A. Harvey, and A. Strominger,
``Quantum black holes,'' UCSB/Chicago preprint UCSBTH-92-08=EFI-92-16,
hep-th@xxx/\-9203042.}
\lref\rom{R. Jackiw, ``Liouville field theory: a two-dimensional model for
gravity?,'' in {\sl Quantum theory of gravity}, S. Christensen, ed.
(Hilger, Bristol U.K. 1984)}
\lref\tei{C. Teitelboim `` The hamiltonian structure of spacetime and its
relation with the conformal anomaly,'' in {\sl Quantum theory of gravity}, S.
Christensen, ed.
(Hilger, Bristol U.K. 1984)}
\Title{\vbox{\baselineskip12pt\hbox{UCSBTH-92-28}
\hbox{hepth@xxx/9207034}
}}
{\vbox{\centerline {Quantum Theories of Dilaton Gravity}
}}

\centerline{{\ticp Steven B. Giddings}\footnote{$^\dagger$}
{Email addresses:
giddings@denali.physics.ucsb.edu, steve@voodoo.bitnet.}
{\ticp and Andrew Strominger}\footnote{$^*$}
{Email addresses:
andy@denali.physics.ucsb.edu, andy@voodoo.bitnet.}
}

\vskip.1in
\centerline{\sl Department of Physics}
\centerline{\sl University of California}
\centerline{\sl Santa Barbara, CA 93106-9530}

\bigskip
\centerline{\bf Abstract}

Quantization of two-dimensional dilaton gravity coupled to conformal matter
is investigated.
Working in conformal gauge about a fixed background metric,
the theory may be viewed as a
sigma model
whose target space is parameterized by the dilaton $\phi$ and conformal factor
$\rho$.  A precise connection is given between the constraint that the
theory be independent of the background metric and conformal invariance of
the resulting sigma model.
Although the action is
renormalizable, new coupling constants must be specified at each order in
perturbation theory in order to determine the
quantum theory.  These constants may be viewed as initial data for the beta
function equations.
It is argued that not all choices of this data correspond to physically
sensible theories of gravity,
and
physically motivated constraints on the data are discussed. In particular a
recently
constructed subclass of initial data which reduces the full quantum theory
to a soluble
Liouville-like theory has energies
unbounded from below and thus is unphysical.
Possibilities for modifying this
construction so as to avoid this difficulty are briefly discussed.

\Date{}

\newsec{Introduction}
Two-dimensional dilaton gravity is a useful model for developing
understanding of the quantum properties of higher-dimensional gravity.
This theory captures several of the essential features of its
higher-dimensional cousins, and in particular has black hole
solutions\refs{\Mandal\WittTwod-\CGHS} and Hawking radiation.   Thus we
might expect it to help us unravel some of the mysteries of real black holes.
In \refs{\CGHS} a program to investigate the collapse and evaporation of
two-dimensional black holes was initiated.  It was shown that effects due
to Hawking radiation of $N$ massless scalar fields are incorporated by
including the Polyakov-Liouville term, arising from the one-loop matter
functional integral, in the action.  A semiclassical $1/N$ treatment of the
resulting equations indeed produces an evaporating black hole.  However,
as was shown in \refs{\BDDO,\RST}, the large-$N$ semiclassical equations become
singular inside the black hole and predictability fails.
The resulting breakdown of the
semiclassical limit necessitates consideration of higher order quantum
corrections to the theory. In \refs{\Stro} the one-loop
semiclassical equations (including effects of the ghost measure) were analyzed
for finite $N$, and such singularities were not found.
However, the interesting physics occurs in a region where the one-loop
approximation can not be justified by
a small expansion parameter. Clearly, it is important to understand the quantum
theory beyond the
one-loop semiclassical level.

In this note we will discuss some aspects of the quantization of  dilaton
gravity.  Much of the discussion is not new: previous work
on this subject can be found in
\refs{\tei\rom\tbo\hver\RuTs\BuCh\deAli\BiCa\deAlii\mik-\ham}. An
important point is that there are infinitely many
theories of quantum dilaton gravity; although the theory is power counting
renormalizable, it is non-predictable in that an infinite
number of coupling constants must be specified to determine the theory, a
finite number of which arise at each order in perturbation theory.
Such  theories can be characterized in terms of initial value
data for the renormalization group equations. However not all choices of
initial
data correspond to physically sensible theories.
For example, recently it was shown that a  subset of the space of
dilaton gravity theories are equivalent to soluble Liouville-like conformal
field theories\refs{\RuTs,\deAli-\deAlii,\GiStunpub}.  We show that this
particular subset does not have physical behavior:
the ADM mass is unbounded from
below\foot{This objection is not relevant to the recent work of
Russo, Susskind and Thorlacius\refs{\rst}, who use the transformation
to Liouville theory as a trick to solve
semiclassical equations, and
are not trying to define the full quantum theory.}. It would be of great
interest to find modifications of this construction which do not have
this difficulty. Constraints on the initial
data required by physical sensibility are
discussed.

\newsec{ Semiclassical Dilaton Gravity Coupled to Conformal Matter}

Dilaton gravity coupled to $N$ conformal matter fields is given by the
action $S_D + S_M$ with
\eqn\sdila{S_D = \frac{1}{2\pi}
\int d^2 x\, \sqrt{-g}\ e^{-2\phi} \left[R+4
(\nabla\phi)^2 + 4\lambda^2\right]}
\eqn\smatt{S_M = - \frac{1}{4\pi}
 \int d^2 x\, \sqrt{-g} \sum^N_{i=1} (\nabla f_i)^2\ .}
At the classical level the theory is exactly soluble \refs\CGHS.
This is easily seen by passing to conformal gauge,
\eqn\twothree{ds^2 = -e^{2\rho} d\sigma^+d\sigma^-,}
with $\sigma^\pm = \sigma^0 \pm \sigma^1$.  The action then becomes
\eqn\lcact{S=\frac{1}{\pi} \int d^2 \sigma\left[-2\partial_+ e^{-2\phi}
\partial_-(\rho-\phi) + \lambda^2 e^{2(\rho-\phi)} + \half \sum^N_{i=1}
\partial_+f_i\pmi f_i \right]\ .}
Given an arbitrary solution $f=f_+(\sigma^+) + f_-(\sigma^-)$ of the matter
equations, the general solution for $\rho$ and $\phi$ is
\eqn\twofive{\eqalign{e^{-2\phi} & = \frac{M}{\lambda} - \lambda^2 \int
d\sigma^+ e^{w_+} \int d\sigma^- e^{w_-}\cr
                                 & -\half \int d\sigma^+ e^{w_+} \int d\sigma^+
e^{-w_+} (\partial_+f)^2 - \half \int d\sigma^- e^{w_-} \int d\sigma^- e^{-w_-}
(\partial_- f)^2\cr
                          \rho-\phi & = \frac{w_+ + w_-}{2}\cr}}
where $w^+(\sigma^+)$ and $w^-(\sigma^-)$ are arbitrary gauge functions.  This
solution includes the general black hole formed by collapse of $f$-matter.

Quantum effects of matter can be understood from the functional integral,
\eqn\twosix{Z= \int {\cal D}(g,\phi,f)
e^{iS_D + iS_M}\ .}
The measure for the matter fields is induced from the natural metric for
matter fluctuations:
\eqn\mmet{<\delta f, \delta f> =\int d^2 x \sqrt{-g} \delta f^2}
and depends on $g$.
The matter functional integral
can be performed and yields
\eqn\twoseven{\int {\cal D} f\ e^{iS_M} = e^{iNS_{PL}}}
with
\eqn\twoeight{S_{PL} = - \frac{1}{96\pi} \int d^2x\ \sqrt{-g(x)}
\int d^2x^\prime \sqrt{-g(x^\prime)}\ R(x) \sq^{-1} (x,x^\prime)
R(x^\prime)}
the Polyakov-Liouville action.  This action incorporates both the
Hawking radiation and its backreaction on the
geometry\refs{\CrFu,\CGHS,\GiNe}.

The semiclassical equations resulting from the large-$N$ action $S_D +
S_{PL}$ have been analyzed in \refs{\CGHS-\RST,\Hawk\SuTh-\BGHS} and
have evaporating black holes as solutions. However, these equations
become singular at a critical value of $\phi$, $e^{2\phi_c} =
\frac{12}{N}$, signalling the need to go beyond a semiclassical
analysis.

\newsec{ Quantum Dilaton Gravity}

Quantization of dilaton gravity proceeds from the functional integral
\eqn\dint{\int {\cal D}( g\, ,\phi)\ e^{iS_D + iNS_{PL}}
\ .}
from which the matter fields have been integrated out.
A convenient approach to gauge fixing and quantization is to express the
metric as a conformal rescaling of a fixed background metric $\hat g$,
\foot{In this topologically trivial context there are no moduli.}
\eqn\bgauge{g=e^{2\rho} \hat g\ .}
%
%
The measure in \dint\ is defined from the metric on the tangent space to
the space of fields\foot{We are grateful to Herman Verlinde for
discussions on this point.}. It is natural to derive this metric from
the metric appearing in the kinetic term in the action \lcact.
In the gauge \bgauge\ this leads to
\eqn\fmet{\eqalign{<\delta\phi,\delta\phi>&=-\int d^2 x \sqrt{-\hat
g}e^{2\rho-2\phi}\delta\phi^2, \cr
<\delta\phi,\delta\rho>&=\int d^2 x \sqrt{-\hat
g}e^{2\rho-2\phi}\delta\phi\delta\rho, \cr
<\delta\rho,\delta\rho>&=0. \cr}}
with a similar metric for the ghosts.
With the gauge and measure choice \bgauge\ and \fmet\ the functional integral
becomes\Stro
\eqn\twotwelve{\int {\cal D}_{e^{2\rho-2\phi}\hat g}(\rho,\phi)\
\det\nolimits_{e^{2\rho-2\phi}\hat g} P\ e^{iS_D + iNS_{PL}}}
where $\det\nolimits_{e^{2\rho-2\phi}\hat g} P$
is the Fadeev-Popov determinant, and the subscripts on
$\cal D$ and det indicate the metric used to define them.  The
naive Ansatz for the transformation of the measures is
\eqn\twothirteen{{\cal D}_{e^{2\rho-2\phi}\hat g} (\rho, \phi)\
\det\nolimits_{e^{2\rho-2\phi}\hat g} P = e^{-24iS_L[\hat g, \rho-\phi]}
{\cal D}_{\hat g}(\rho, \phi) \det\nolimits_{\hat g} P}
with the Liouville action
\eqn\liou{S_L[\hat g, \rho-\phi] =  \frac{1}{24\pi} \int d^2 x\, \sqrt{-\hat
g}\ \left(\widehat{(\nabla(\rho-\phi))^2} + \widehat R(\rho-\phi)\right)\ .}

Note that the $\rho,\phi$ dependence of the
ghost-gravity measure differs from that of the
matter measure. We have motivated that difference here by
choosing the measure naturally associated with the action \sdila.
This choice is somewhat arbitrary in that other choices of measure (e.g.
obtained by
shifting $\rho$ by different multiples of $\phi$)
would also lead to covariant theories, corresponding to the freedom of adding
finite, covariant, local counterterms at the one-loop level. However
the physically sensibility of the choice we have made is
confirmed by the observation \Stro\ that it implies, at one-loop
order, that black holes do not Hawking radiate ghosts or (non-dynamical)
$\rho$ or $\phi$ modes.
%

The ambiguity associated with finite counterterms does not end at the one-loop
level.
Although dilaton gravity is renormalizable,
this doesn't buy as much in two dimensions as in higher dimensions since
the fields $\rho$ and $\phi$ have dimension zero.  In general,
quantization will therefore introduce renormalizable counterterms of the
form\foot{One might also consider the addition of $(1,1)$ operators with more
than
two derivatives, corresponding to massive string modes. We will ignore this
possibility in the
following.}
\eqn\gcount{S=-{1 \over2 \pi}\int d^2 x\ \sqrt{-\hat g}
\left(G_{\mu\nu}(X^\lambda)
\widehat{\nabla X^\mu \cdot \nabla X^\nu} +
\half \Phi (X^\lambda) \hat R + T(X^\lambda)
\right)\ ,}
with $X^\lambda = (\rho, \phi)$ (or more general coordinates on $\rho,\phi$
space). In this sense, dilaton gravity is
closer to its non-renormalizable cousins in higher dimensions than we
might have hoped.  Indeed, to define the theory we must specify an
infinite number of coupling constants (giving the complete functions in
\gcount).  Therefore the theory is non-predictable: a two-dimensional
observer must perform an infinite number of experiments to determine the
lagrangian.

In quantum dilaton gravity, one must therefore consider
$\sigma$-model actions of the form \gcount.  However, as
discussed in several papers \refs{\jptb} , not every action of the
form \gcount\ corresponds to a theory of gravity.
One one way of stating the reason for this is two-dimensional general
covariance, or equivalently background independence:
given the decomposition \bgauge, the theory should be invariant under the
transformation
\eqn\bgind{\hat g \rightarrow e^{2\delta\omega} \hat g\ ,\ \rho\rightarrow
\rho-\delta\omega}
(quantum corrections may modify the form of the latter transformation).
This is equivalent to the requirement that
\gcount\ be a $c=26-N$ conformal field theory.
To see the reason for this
this connection, note that quantization of the theory by fixing the gauge
as in \bgauge\ leaves unfixed
a group of residual diffeomorphisms which is isomorphic to the conformal
group. These residual symmetries
are generated by a set of operators in the theory which
obey (two copies of ) the Virasoro algebra (with $c=0$ when ghosts and matter
are included).
The correlation functions
will then exhibit the corresponding Ward identities,
which may be taken as the
defining
characteristic of a conformal field theory.

Thus the couplings
$G,~\Phi$ and $T$ are restricted by conformal invariance
to satisfy $\beta$-function equations.
Except in very special cases it is not possible to solve these equations
exactly; one must find a perturbation expansion. One such expansion is the
loop expansion\foot{This is related to the ${\alpha^{\prime}}$
expansion in string theory, but it is not identical because in string theory
different powers
of ${\alpha^{\prime}}$ conventionally appear
in front of the three terms in
\gcount. That reordering of the loop expansion is convenient in string theory
but not in the present context.} in $\hbar$.
{}From \sdila\ it is evident that an equivalent loop
expansion parameter for dilaton gravity is ${e^{2\phi}}$, so this
will be useful when ${e^{2\phi}}$ is small.  Also by working in the
loop expansion, we can provide a more explicit description of the
connection between
background independence and conformal invariance.

At the classical level, the conditions on ${G}, {\Phi}$ and ${T}$ that
follow from invariance under the transformation \bgind\
should be precisely those which enable one
to eliminate ${\hat g}$ in \gcount\ and rewrite it in a covariant form
as a theory of dilaton gravity. We will now verify that this is indeed the
case, and find the connection to conformal invariance.
Let us consider the generalization of the transformation \bgind\
incorporating
some (as-yet-unspecified)
variation ${\delta}{X^{\mu}}$ of the ${X^{\mu}}$'s.  Invariance of the
action requires
\eqn\cvr
{\eqalign{0&={2\pi\over\sqrt{-\ghat}}
({{{\delta}{S}}\over{{\delta}{\hat g}}}
{\delta{\hat g}} + {{{\delta}{S}}\over{{\delta}{X^\mu}}}
{\delta}{X^\mu})\cr
&=({\partial_\mu}{\partial_\nu}{\Phi}\hat{\nabla X^\mu\cdot\nabla X^\nu} +
{\partial_\mu}{\Phi}{\hat\sq}{X^\mu} - {2T}){\delta\omega}\cr
&\quad + {2}{\pi} {{{\delta}{S}}\over{{\delta}{X^\mu}}} {\delta}{X^\mu}.\cr}}
where
\eqn\eom{
{{2\pi}\over\sqrt{-\ghat}} {{{\delta}{S}}\over{{\delta}{X^\mu}}} =
2G_{\mu \lambda}{\hat\sq}
{X^\lambda}
+ 2{\Gamma}_{\mu,\nu\lambda}\hat{\nabla X^\nu\cdot\nabla X^\lambda}
{-} {{1}\over{2}} {\partial_\mu} {\Phi}{\hat R} -
{\partial_\mu}{T}
}
and spacetime indices are suppressed.
Eq.~\cvr\ may be rewritten in a target-space covariant form:
\eqn\rxp
{\eqalign{0&=({\beta_{0\mu\nu}^G}\hat{\nabla X^\mu\cdot\nabla {X^\nu}}
+ {{1}\over{2}} {\beta_0^\Phi} {\hat R} + {\beta_0^T})
{\delta\omega}\cr
&+ {2\pi\over\sqrt{-\ghat}}
{{{\delta}{S}}\over{{\delta}{X^\mu}}} ({\delta}{X^\mu} +\half
{\nabla^\mu}{\Phi}{\delta\omega}),\cr}}
\noindent where the zeroth-order beta functions are
\eqn\bfn
{\eqalign{{\beta_{0\mu\nu}^G}&={\nabla_\mu}{\nabla_\nu}{\Phi},\cr
{\beta_0^\Phi}&={{1}\over{2}} ({\nabla}{\Phi})^2,\cr
{\beta_0^T}&={{1}\over{2}}{\nabla_\mu}{\Phi}{\nabla^\mu}{T} {-} {2T}.\cr}}
One sees that (to leading order)
background independence and conformal invariance
are equivalent on-shell.  More generally, \rxp\ shows that the theory is
background independent even off-shell
if the $\beta$-functions vanish,
\eqn\bvn
{{\beta_0^G} = {\beta_0^\Phi} = {\beta_0^T} = {0},}
and if the variation $\delta X^\mu$ is given by
\eqn\xvr
{{\delta}{X^\mu} = -\half{\nabla^\mu}{\Phi}{\delta\omega}.}

The vanishing of the ${\beta}$'s can be interpreted geometrically:
${\beta_0^G} = {0}$ implies that ${k_\nu} = {\nabla_\nu}{\Phi}$ is
a Killing vector, while ${\beta^\Phi} = {0}$ implies ${k}$ is null. Every
two-dimensional geometry with a null Killing vector is flat. If the Killing
vector
is the gradient of a globally defined function, the space is flat
two-dimensional Minkowski space
up to possible periodic identifications in the direction
transverse to $k$. It also follows that $\Phi$ runs from $+\infty$ to $-\infty$
as $X$ runs over the target space.

In order to check that the general background-independent lagrangian \gcount\
can be rewritten as a theory of dilaton gravity it is necessary
to specify ${\rho}$ and ${\phi}$ in terms of the general fields ${X^\lambda}$.
This identification is determined from \xvr\ which we would like to identify
with the Weyl transformation as in \bgind.
Under Weyl transformations one should find ${\delta}{\phi} = {0}$ while
${\delta\rho} = -{\delta\omega}$. The correct transformation law for
${\phi}$ is obtained by defining
\eqn\zzz
{{\phi} = {\phi} ({\Phi}({X}))}
\noindent for a smooth single valued function ${\phi}({\Phi})$
(for example $\phi=\Phi$). There is no
preferred function, corresponding to the inherent ambiguity in field
redefinitions of ${\phi}$.  However we shall see below that regularity
of $G$ in the $(\rho, \phi)$ coordinates requires that $\phi$ is
a monotonically increasing function of $\Phi$.
This then implies that the range of
$\phi$ is unrestricted.\foot{Note however that in the action \sdila,
$\Phi=-2e^{-2\phi}$.  If only finite real values of $\phi$ are allowed,
then the sigma-model target space has a boundary at $\Phi=0$.}
One finds using \xvr\ and ${\beta_0^\Phi}
= {0}$ that
\eqn\qqq
{{\delta\phi} = {{\del\phi}\over{\del\Phi}} \nabla_\mu \Phi \delta X^\mu =
-\half{\del\phi\over\del\Phi}
({\nabla}{\Phi})^2
\delta \omega = {0},}
\noindent as desired.

To define ${\rho}$ as a function of the ${X^\mu}$'s, note that the flatness
of ${G}$ implies the existence of a second Killing vector ${\ell_\mu}$
obeying
\eqn\lkl
{{\nabla}_{\mu}{\ell}_{\nu} = {0}}
\noindent and
\eqn\thg
{{\ell^\mu}{k_\mu} = {1}.}
The conformal factor ${\rho}$ may then be defined by
\eqn\rdf
{{\rho}({X}) = 2{\int^X}{\ell_\mu}{d}{X^\mu}.}
\noindent The variation of ${\rho}$ is then
\eqn\rvr
{\eqalign{
{\delta\rho} = 2{\ell_\mu}{\delta}{X^\mu}&=
-{\ell_\mu}{\nabla^\mu}{\Phi}{\delta
\omega}\cr
&=-{\delta\omega}.\cr}}
in agreement with \bgind. Note that single-valuedness of
$\rho$ requires that there are $not$ periodic identifications of $X$ as
discussed below equation \xvr. The range of $\rho$ is then unrestricted.
There is also an ambiguity in ${\rho}$ corresponding
to shifts by arbitrary functions of ${\phi}$.

Constraints may now be obtained on the components of ${G}$ in the
${\rho},{\phi}$ coordinates.
Firstly, using $({\ell}_\rho, \ell_\phi) =(\half,0)$ and
$(k_\rho,k_\phi) = (0,{{\del\Phi}\over{\del\phi}})$, \thg\ becomes
\eqn\grf
{{G}^{\rho\phi} = 2{{\del\phi}\over{\del\Phi}}.}
\noindent Secondly ${\beta_0^\Phi} = {0}$ implies
\eqn\gff
{{0} = \biggl({{\del\Phi}\over{\del\phi}}\biggr)^2 {G}^{\phi\phi}}
\noindent or equivalently
\eqn\grr
{{G}_{\rho\rho} = {0}.}
\noindent Finally, given \grr\ and \grf\  flatness of ${G}$ requires
that
\eqn\gff
{{G}_{\phi\phi} = {K}({\phi})}
\noindent and is independent of ${\rho}$.

The equation ${\beta_0^T} = {0}$ for ${T}$ is, in ${\rho,\phi}$ coordinates
simply
\eqn\trf
{{{\partial T}\over {\partial\rho}} = {2T}.}
\noindent The general solution to this is
\eqn\tsl
{{T} = {e^{2\rho}}{U}({\phi}),}
\noindent for some ${\rho}$-independent function ${U}$.

Using \grr\ , \grf\ , \gff\ and \trf\ , \gcount\ may be rewritten
in ${\rho,\phi}$ coordinates as
\eqn\snt
{\eqalign{
{S}&= -{{1}\over{2\pi}}\,\,{\int}{d^2}{x}{\sqrt{{{-}{\hat g}}}}\,\,({K}
\hat{\nabla\phi\cdot\nabla} {\phi}\cr
&+{{\partial \Phi}\over{\partial \phi}} \hat{\nabla\phi\cdot\nabla} \rho
+ {{1}\over{2}} {\Phi}{\hat R} + {e^{2\rho}} {U}).\cr}}
\noindent Defining
\eqn\ggh
{{g} = {e^{2\rho}}{\hat g}}
\noindent and integrating by parts one finds
\eqn\sfn
{{S} = -{{1}\over{2\pi}}\,\,{\int} {d^2}{x} {\sqrt{-g}}
\left({K}{\nabla}{\phi}{\cdot}{\nabla}{\phi}+
{{1}\over{2}}{\Phi}{R}   + {U}\right).}

Eq.~\sfn\ is the most general power-counting renormalizable theory of
two-dimensional gravity coupled to a single scalar field  ${\phi}$.
Thus we see explicitly that classical background independence or conformal
invariance of the
sigma
model of the form \gcount\ implies equivalence
to a theory of gravity. Conversely,
by repeating the preceding steps in reverse, it can be seen that every
theory of gravity coupled to a scalar field is equivalent to a
conformally invariant sigma model of the form \gcount.

As discussed above, this equivalence persists at the quantum level. Quantum
conformal invariance requires that the quantum corrected beta functions
vanish.
To first order in the $\hbar$ expansion these take the
form (temporarily reinstating $\hbar$)
%
\eqn\bfuns{\eqalign{
 \beta^T  & =
\half \nabla_\mu\Phi \nabla^\mu T -2T -{\hbar\over4}\sq
T+\cdots =0,\cr
\beta^G_{\mu\nu} & =
\nabla_\mu\nabla_\nu \Phi + {\hbar\over2} R_{\mu\nu}+\cdots=0,\cr
\beta^\Phi& =\half
\left( \nabla \Phi\right)^2 -{\hbar\over4} \sq\Phi+ 2\hbar\gamma
+\cdots=0,}}
where ${\cal R}$ is the curvature of $G$ and we define
\eqn\gmm{\gamma \equiv {N-24 \over 12}.}

There are many solutions of the $\beta$-function equations. This may be
viewed as a target space initial value problem. For example we could specify
$G,~ \Phi$ and $T$ as functions of $\phi$ at fixed $\rho$, and then
use the $\beta$ function equations to determine them at other values
of $\rho$.\foot{The target space initial data is subject to the usual
constraints of
a covariant theory. These can be used to solve for $G$ and $\Phi$ (up to
a finite number of integration constants), leaving $T$ as the only freely
specifiable function. There are an infinite number
of further functions (corresponding to the massive string modes) in
the most general setting.}
In
general, explicitly writing down these equations, let alone solving them, is a
difficult task.  Nonetheless, we can hope to make progress by
considering special cases where they simplify.

In particular, it is worth pointing out that the
loop expansion is not the only way to investigate solutions of the $\beta$
function equations.  An alternative procedure is to begin with a solution
that is known to be an exact conformal field theory on other grounds, and
then perturb it in the direction defined by some marginal
operator.  In this
case the small expansion parameter is the coefficient
of the perturbation.  For the dilaton gravity theory
\sdila, this expansion is good when the cosmological constant term,
and hence $e^{-2\phi}$, is $small$. This is in contrast to the usual
loop expansion about the linear dilaton vacuum, which is good
when $e^{-2\phi}$ is $large$.
There are thus in general two inequivalent
(and possibly overlapping) expansion
schemes that may be helpful in finding particular solutions.
We will consider one such special case where the
marginal operator expansion can be
used in the next section.

However, even if we could find all solutions to the $\beta$-function
equations, not all of these covariant theories are useful for studying
black hole physics. We are interested in those theories which
reduce to dilaton gravity in the classical ($e^{2\phi}\rightarrow 0$)
limit. The subleading (one-loop) $e^{2\phi}$ dependence is also constrained
by the demand that Hawking radiation should proceed only by matter emission,
and to leading
non-trivial order should take the form described in \refs{\CGHS,\Stro}.
These conditions are enforced by requiring the full action
\gcount\ to duplicate
\eqn\fullact{S_D + NS_{L} \left[\hat g,\, \rho \right] - 24S_L \left[\hat g,\,
\rho-\phi \right]}
to one-loop level at weak coupling.  Rewriting \fullact\ in the form
\gcount, this gives
\eqn\asmet{G_{\mu \nu}
\buildrel \longrightarrow\over{\scriptscriptstyle \phi\rightarrow - \infty}
\left(\matrix{-4e^{-2\phi}+2 & 2e^{-2\phi} - 2\cr
              2e^{-2\phi}-2 & -{\gamma}\cr}\right) + {\cal O}
(e^{2\phi}),}
\eqn\asdil{\Phi \buildrel\longrightarrow\over{\scriptscriptstyle
\phi\rightarrow - \infty}
-2 e^{-2\phi} -4 \phi -{2\gamma }\rho + {\cal O} \left(e^{2\phi}\right),}
and
\eqn\astach{T \buildrel\longrightarrow\over{\scriptscriptstyle
\phi\rightarrow - \infty}
 -4\la^2e^{2\rho-2\phi} + {\cal O} \left(1 \right).}

There are still further conditions which must be met if
the theory is to serve as a useful model for
studying black hole dynamics. One such physical condition is that the
higher order corrections preserve the existence of a stable ground state.

In summary, we wish to study quantum theories of dilaton  gravity
described by actions of the form \gcount, where $G,~\Phi$ and $T$ are
constrained by the requirements that
\item{1.} They must agree with the couplings of classical dilaton gravity to
leading order in $e^{2\phi}$.
\item{2.} Hawking radiation should proceed only by matter emission, and to
leading
non-trivial order the action
should take the form described in \refs{\CGHS,\Stro}.
\item{3.} They should define a $c=26-N$ conformal field theory.
\item{4.}There must be a ground state of the full quantum theory.

Further investigations of the physical sensibility of these models may
lead to yet further restrictions.

\newsec{Soluble Models}

One approach to solving \bfuns, pursued by deAlwis \refs{\deAli,\deAlii} and
Bilal
and Callan\BiCa, is to search for appropriate exact
conformal theories.  In particular, the asymptotic
target space metric \asmet\ is
easily seen to be lorentzian and
flat. They therefore consider the special case of exactly
flat
lorentzian metrics (in this section we take $\hat g$ to be flat
and use  coordinates $\sigma^\pm$)
\eqn\flmet{G_{\mu \nu}\p+ X^\mu \pmi X^\nu =-\gamma\p+X\pmi X
+\frac{1}{\gamma}\p+ Y \pmi Y .}
The relationship of the natural flat coordinates $X$ and $Y$ with $\rho$ and
$\phi$ will be
described momentarily.
To leading order, eqs.~\bfuns\ then
imply
\eqn\fourtwo{\Phi =- 2 \gamma X}
and that the operator corresponding to the tachyon background,
\eqn\fourthree{T= -4\lambda^2 e^{2(X-Y/\gamma)}\ }
is $(1,1)$. In fact, the resulting action
\eqn\ecft{ S= {1 \over \pi}\int d^2 \sigma\,\left[-\gamma\p+X\pmi X
+\frac{1}{\gamma}\p+ Y \pmi Y +\lambda^2
 e^{2( X-\frac{Y}{\gamma})} \right]}
defines an exact CFT, hence an exact solution of \bfuns.

The lagrangian \ecft\ is also classically soluble. The change of
coordinates
\eqn\fourfive{\eqalign{V&=  X-\frac{Y}{\gamma}\cr
                       U&=\half ( \gamma X +Y)\cr}}
puts it in the form
\eqn\foursix{S= {1 \over \pi}\int d^2\sigma \left(-2\p+U\pmi V +\lambda^2
e^{2V}\right)}
which is the same as the gravitational part of \lcact\ if one identifies
\eqn\fourseven{\eqalign{V& \leftrightarrow
\rho-\phi\cr
                        U& \leftrightarrow e^{-2\phi}
\ .\cr}}
This has the general classical solution
\eqn\uvsol{\eqalign{U&= u_+ + u_- - \lambda^2 \int d\sigma^+ e^{w_+} \int
d\sigma^- e^{w_-}\cr
                    V&= \frac{w_+ + w_-}{2}\cr}}
where $u_+$ and $u_-$ are determined by the classical
matter distribution as in \twofive.

$X$ and $Y$ are related to the usual variables $\rho$ and $\phi$ for dilaton
gravity
via a target space coordinate transformation, which restores the action to the
original form  \asmet-\astach\ (up to order $e^{2\phi}$). Let
\eqn\corresp{\eqalign{Y&= - \int d\phi\ \sqrt{4e^{-4\phi} - 4(\gamma +
2) e^{-2\phi} + 2(\gamma+2)}\ +\delta Y\cr
              X&=\rho  + \frac{e^{-2\phi} +
2\phi}{\gamma} + \delta  X\cr}}
where
\eqn\fourten{\delta Y, \delta  X \sim {\cal O}
\left(e^{2\phi}\right)}
and
\eqn\foureleven{\delta \left( X - \frac{Y}{\gamma}\right) \sim
{\cal O} \left( e^{4\phi}\right)\ .}
It is straightforward to verify that the target space
fields $G,~~\Phi$ and $T$ take the form \asmet-\astach\
in terms of the new variables.

This relation can be illustrated by considering the static
solutions.  Taking $w_+ = \la \sigma^+$, $w_-=-\la \sigma^-$, and $u_+ +
u_-=\mu + F\sigma$,
(with $F$ a constant and $\sigma^+-\sigma^- =2 \sigma$) in \uvsol, one finds
\eqn\ysolu{U= \mu + F\sigma + e^{2\la \sigma} }
and
\eqn\rsolu{ V = \la \sigma\ .}
One can solve for the $\sigma \rightarrow \infty$ behavior of $\rho$
and $\phi$ by expanding \corresp\ in  $e^{2\phi}$ and equating the
expansions to \ysolu, \rsolu. One finds for large negative $\phi$
\eqn\uvfr{\eqalign{U &={\gamma \over 2}\rho +e^{-2\phi}+({\gamma \over 2}
+2)\phi+{\gamma (\gamma +2)\over 16}
e^{2\phi}+ ~.~.~.\cr
V &=\rho-\phi-{\gamma+2 \over 8}e^{2\phi}+~.~.~.\cr}}
which then implies
\eqn\fourfourteen{\eqalign{\phi&= -\la \sigma - \half \left[\mu + \left(F+
\frac{\gamma \la }{2} + 2 \la\right) \sigma\right] e^{-2\la \sigma} + {\cal O}
\left(e^{-4\la\sigma}\right)\cr
                          \rho&= -\half \left[\mu - \frac{\gamma +2}{4} +
\left(F + \frac{\gamma\la}{2} + 2\la\right) \sigma\right] e^{-2\la\sigma} +
{\cal
O} \left(e^{-4\la\sigma}\right)\ .\cr}}
For $F+\frac{\gamma\la}{2} + 2\la = 0$ these are asymptotic solutions similar
to the usual black holes, modulo the extra term in $\rho$.  The
ADM mass may be computed in the usual fashion, by linearizing the
constraints about a given solution, \eg\ that with $\mu =
0$. One finds
\eqn\ntysix{
\eqalign{M&=2 e^{\lambda(\sigma^+ - \sigma^-)} (\lambda \delta \rho +
               \partial_+ \delta \phi - \partial_- \delta \phi)\cr}}
where $\delta \rho$ and $\delta \phi$ are the asymptotically vanishing
deviations of $\rho$ and $\phi$ from their $\mu=0$ values.
%
The resulting mass is the expected
\eqn\fourfifteen{M=\la \mu  .}
The solutions with $F+\frac{\gamma}{2\la} + 2\la\not=0$ have infinite mass
\foot{Although a finite mass $M'$ can be defined
for solutions with any value of $F$
by linearizing around a reference solution with that fixed value of $F$.
$M'$ is then easily seen to be invariant under asymptotically trivial field
redefinitions, which explains why $M$ in \fourfifteen\ agrees with the mass
defined directly in the $U-V$ variables from \ysolu\ and \rsolu.},
like the black holes in equilibrium with incoming radiation described in
\refs\BGHS.

At first sight it appears that one has made great progress by finding an
exact conformal field theory describing black holes together with
Hawking radiation.  However, further examination reveals a difficulty with
the identification of the Liouville-like model, \ecft, as a physical theory
of dilaton gravity.
In the conformal field theory variables $U$ and
$V$, the solution \ysolu, \rsolu\ is non-singular independent of the
value of $\mu$, and for all $\sigma$: $U=0$ is a regular value for the
field.  Thus, in particular, there is nothing wrong with taking $\mu$
large and negative\foot{In fact the theory we wish to describe presumably
does not have static solutions
with a continuously varying mass of any sign:
black holes should evaporate.}. This means that the mass $M$ is unbounded from
below, and the theory doesn't have a ground state, in contradiction with
our requirement (4.)\foot{This is in agreement with the observations of
\refs{\BiCa,\deAlii}
that the Hawking radiation rate asymptotes to a constant value in the future,
rather than shutting off when
the mass reaches zero. The conclusion here is not
affected by quantum corrections to the mass formula discussed in \past.}.
Interpretation of
the theory \ecft\ as a physical model for black hole physics is
therefore very problematic.
One might
attempt to evade this problem by constraining the range of $U$ and $V$ in such
a way that the
mass is bounded. However this is throwing out the baby with the bath water: the
resulting constrained theory is
very complicated and is unlikely to correspond to a soluble conformal field
theory.

\newsec{Conclusions}

We have argued that there are an infinite number of quantum theories of
dilaton gravity.  These are parameterized by initial value data for the
renormalization group equations \bfuns\ that produce solutions with the
asymptotic behavior \asmet, \asdil.  One would like to find further
criteria to narrow the class of solutions, and a more
explicit description of the allowed solutions. One approach is to
specify initial data corresponding to a known soluble conformal field theory,
but we have found that the
simplest such choice leads to a theory with unphysical behavior.

Having said this, we would like to add that we nevertheless find the basic idea
of
deAlwis and Bilal and Callan of fixing the higher order corrections in order to
obtain
a soluble conformal field theory
very appealing. If this were indeed possible, it would certainly greatly
clarify the
quantum properties of two-dimensional black holes.
While their attempt was not completely successful, perhaps some variant on this
idea may yet
work. For example there are, in addition to \fourthree, a one parameter family
of solutions of
the tachyon equation of motion which correspond to $(1,1)$ operators. Some of
these vanish
at weak coupling, and could be used to perturb the lagrangian. Alternately, the
target
space metric need not be flat beyond leading order. Thus one might consider a
conformal theory
with an asymptotically flat but curved target space, such as the $SU(1,1)/U(1)$
black hole
(perturbed by a tachyon), as
a candidate quantum extension of classical dilaton gravity\foot{Much more
generally, studies of (tree level) $c=1$ matrix models have produced (albeit in
an
unusual and indirect manner) infinite families of exact conformal field
theories
corresponding to a metric, dilaton and tachyon in a two-dimensional target
space.
Perhaps this spectacular technological development has applications in the
present context.}.

On the other hand
there is no obvious a priori reason to believe that a physical theory
describing black hole evaporation should correspond to a simple
conformal field theory. (Indeed the existence of a unique static solution
suggests the opposite.)  In general, one must confront the
difficult problems of finding physically sensible initial data and
solving the renormalization group equations \bfuns.
One approach is to try to generally characterize the space of physically
sensible theories,
together with the behavior of black holes in such theories. Another approach is
to
try find and analyze special examples of such theories. Such special examples
may be
obtained from
low-energy limit of two-dimensional string theory, or extremal black holes in
higher-dimensional
string theory. Alternately, extended supersymmetry might be used to constrain
the
space of theories. All of these paths
deserve further exploration.

\bigskip\bigskip\centerline{{\bf Acknowledgements}}\nobreak
This work was supported in part by DOE grant DOE-91ER40618 and
by NSF PYI grant PHY-9157463 to SBG.

\listrefs

\end